# АЛГОРИТМ ОЦЕНИВАНИЯ ПЕРЕМЕННОЙ ЧАСТОТЫ СИНУСОИДАЛЬНОГО СИГНАЛА

С. И. Низовцев, С. В. Шаветов, А. А. Пыркин

*Университет ИТМО, 197101, Санкт-Петербург, Кронверкский пр., д. 49, лит. А, Российская Федерация, s.shavetov@itmo.ru*

В статье рассматривается задача идентификации переменной частоты сигнала синусоидальной формы. Для получения регрессионной модели сигнала выполняется итеративное дифференцирование исходного аналитического выражения и применяется лемма о перестановках. Оценивание параметров нестационарной частоты реализовано с помощью процедуры динамического расширения регрессора и смешивания (DREM) и наблюдателя Люенбергера. В результате проведенного численного моделирования продемонстрирована работоспособность предложенного алгоритма, показывая сходимость оценки частоты к истинному значению.

**Ключевые слова:** идентификация параметров, нестационарная частота, наблюдатель Люенбергера, метод динамического расширения и смешивания регрессора.

## Введение

Задача идентификации частоты периодического сигнала крайне актуальна в большом количестве практических приложений: системы динамического позиционирования и компенсации возмущений, системы виброзащиты, системы мониторинга при определении параметров высотных или большепролетных строительных сооружений [1]. Как правило, такие задачи решаются с использованием быстрого преобразования Фурье при накоплении измерений колебаний в течение длительного промежутка времени (от одного часа). В статье рассматривается подход, позволяющий кратно снизить необходимое время наблюдения, и позволяющий оценивать параметры периодических сигналов за достаточно короткий промежуток времени.

В теории идентификации [2] широко распространен подход, когда модель динамической системы приводится к линейной регрессионной модели, см. например [3-5]. В настоящей статье также используется данный подход, однако важным отличием является рассмотрение математической модели, которая содержит неизвестный нестационарный параметр. В публикациях [6, 7] используется алгоритм DREM для замены регрессионной модели $n$-го порядка скалярными регрессиями, после чего оцениваются параметры отдельно с использованием метода градиентного спуска. В статье [8] используется алгоритм DREM для параметрической оценки затухающих синусоидальных сигналов.

Известен результат [9], где при допущении, что нестационарный параметр является произведением постоянного коэффициента на измеряемую функцию времени, было синтезировано устройство оценки для линейной системы, с подобного рода нестационарными параметрами. Решение аналогичной задачи для линейных нестационарных систем представлено в работах [10-13]. Случай, когда на конечных интервалах времени изменяются во времени параметры по линейному закону с постоянным временем, представлен в публикациях [10, 11]. Решение задачи синтеза наблюдателя для нестационарной системы с полиномиальными параметрами произвольной степени опубликовано в статьях [12, 13]. Основным недостатком упоминаемых работ является неочевидность применения для реальных практических задач.

Представленный в статье алгоритм может использоваться в системах мониторинга состояния зданий [14, 15] и существенно расширить сферы применения данных систем, например, на задачи мониторинга состояния секций высотных металлоконструкций при их монтаже, что затруднительно реализовать с использованием известных методов.

**Постановка задачи**

Рассматривается модель гармонического сигнала $y(t)$ следующего вида:
$$y(t) = A \sin(\omega(t) \cdot t + \varphi), \qquad (1)$$

где $A$ – постоянная амплитуда, $\varphi$ – фаза, $t$ – время, $\omega(t)$ – нестационарная частота такая, что:
$$\dot{\omega}(t) = \beta_i, \qquad (2)$$

где $\beta_i = \text{const}$ для $\forall t \in [t_i, t_{i+1})$.

Требуется синтезировать алгоритм оценивания неизвестной частоты $\hat{\omega}(t)$:
$$\lim_{t \to \infty}(\omega(t) - \hat{\omega}(t)) = 0.$$

**Параметризация модели**

Представим гармонический сигнал (1) в виде регрессионной модели:
$$Z(t) = \Theta(t)^{\mathrm{T}} \Psi(t), \qquad (3)$$

где $\Theta(t) = \begin{bmatrix} f(\beta) \\ f(\omega) \\ f(\beta, \omega) \end{bmatrix}$ – регрессор, сигналы $y(t), Z(t), \Psi(t)$ – измеряются. Для удобства дальнейшего использования опустим аргумент $t$: $y = y(t)$, $\omega = \omega(t)$.

Продифференцируем $y$ трижды:
$$\dot{y} = A \cos(\omega t + \varphi)(\omega + \beta t),$$
$$\ddot{y} = -A \sin(\omega t + \varphi)(\omega + \beta t)^2 + 2A \cos(\omega t + \varphi),$$
$$\dddot{y} = -A \cos(\omega t + \varphi)(\omega + \beta t)^3 - A \sin(\omega t + \varphi)\bigl(6\beta(\omega + \beta t)\bigr).$$

С учётом соотношений (1) и (2) выразим $\dddot{y}$ через $y$ и $\dot{y}$:
$$\dddot{y} = -\beta^2 t^2 \dot{y} - 6\beta^2 t y - 2\beta \omega t \dot{y} - 6\beta \omega y - \omega^2 \dot{y}. \qquad (4)$$

Применим фильтр $\dfrac{\lambda^3}{(p+\lambda)^3}$ к выражению (4) и представим его в виде следующего набора фильтров (см. подробные вычисления в Приложении 1):

$$\frac{\lambda^3 p^3}{(p+\lambda)^3} y = -\beta^2 \frac{\lambda^2}{(p+\lambda)^2}\left[t^2 \frac{\lambda p}{(p+\lambda)}[y]\right] + 2\beta^2 \frac{\lambda^2}{(p+\lambda)^3}\left[t \frac{\lambda p}{(p+\lambda)}[y]\right] +$$
$$+ 6\beta^2 \frac{1}{(p+\lambda)}\left[t \frac{\lambda^3 p}{(p+\lambda)^3}[y]\right] - 6\beta^2 \frac{\lambda^3}{(p+\lambda)^3}[ty] - 6\beta\omega \frac{\lambda^3}{(p+\lambda)^3}[y] + \qquad (5)$$
$$+ 18\beta^2 \frac{\lambda^3}{(p+\lambda)^4}[y] - (2\beta\omega t + \omega^2)\frac{\lambda^3 p}{(p+\lambda)^3}[y] + 12\beta\omega \frac{\lambda^3 p}{(p+\lambda)^4}[y] -$$

$$-30\beta^2 \frac{\lambda^3 p}{(p+\lambda)^5}[y].$$

## Алгоритм оценивания частоты

Уравнение (5) с учетом выражения (3) представим в следующем виде:

$$Z(\lambda) = \beta^2 \cdot \Psi_1(\lambda) + \beta\omega \cdot \Psi_2(\lambda) + \omega^2 \cdot \Psi_3(\lambda), \qquad (6)$$

где

$$Z(\lambda) = \frac{\lambda^3 p^3}{(p+\lambda)^3} y, \qquad (7)$$

$$\Psi_1(\lambda) = -\frac{\lambda^2}{(p+\lambda)^2}\left[t^2 \frac{\lambda p}{(p+\lambda)}[y]\right] + 2\frac{\lambda^2}{(p+\lambda)^3}\left[t \frac{\lambda p}{(p+\lambda)}[y]\right] +$$

$$+6\frac{1}{(p+\lambda)}\left[t \frac{\lambda^3 p}{(p+\lambda)^3}[y]\right] - 6\frac{\lambda^3}{(p+\lambda)^3}[ty] + 18\frac{\lambda^3}{(p+\lambda)^4}[y], \qquad (8)$$

$$\Psi_2(\lambda) = -6\frac{\lambda^3}{(p+\lambda)^3}[y] - 2t\frac{\lambda^3 p}{(p+\lambda)^3}[y] + 12\frac{\lambda^3 p}{(p+\lambda)^4}[y], \qquad (9)$$

$$\Psi_3(\lambda) = -\frac{\lambda^3 p}{(p+\lambda)^3}[y]. \qquad (10)$$

Применим процедуру динамического расширения регрессора и смешивания DREM. Для этого подставим три параметра $\lambda_1, \lambda_2, \lambda_3$ в выражения (7)-(10):

$$\lambda_1: Z(\lambda_1) = \beta^2 \cdot \Psi_1(\lambda_1) + \beta\omega \cdot \Psi_2(\lambda_1) + \omega^2 \cdot \Psi_3(\lambda_1), \qquad (11)$$

$$\lambda_2: Z(\lambda_2) = \beta^2 \cdot \Psi_1(\lambda_2) + \beta\omega \cdot \Psi_2(\lambda_2) + \omega^2 \cdot \Psi_3(\lambda_2), \qquad (12)$$

$$\lambda_3: Z(\lambda_2) = \beta^2 \cdot \Psi_1(\lambda_3) + \beta\omega \cdot \Psi_2(\lambda_3) + \omega^2 \cdot \Psi_3(\lambda_3). \qquad (13)$$

Представим в матричном виде уравнения (11)-(13):

$$\begin{bmatrix} Z_1 \\ Z_2 \\ Z_3 \end{bmatrix} = \begin{bmatrix} \Psi_{11} & \Psi_{12} & \Psi_{13} \\ \Psi_{21} & \Psi_{22} & \Psi_{23} \\ \Psi_{31} & \Psi_{32} & \Psi_{33} \end{bmatrix} \begin{bmatrix} \beta^2 \\ \beta\omega \\ \omega^2 \end{bmatrix}. \qquad (14)$$

В выражении (14) введем обозначения:

$$Z = \begin{bmatrix} Z_1 \\ Z_2 \\ Z_3 \end{bmatrix}, \Psi = \begin{bmatrix} \Psi_{11} & \Psi_{12} & \Psi_{13} \\ \Psi_{21} & \Psi_{22} & \Psi_{23} \\ \Psi_{31} & \Psi_{32} & \Psi_{33} \end{bmatrix}, \Sigma = \begin{bmatrix} \beta^2 \\ \beta\omega \\ \omega^2 \end{bmatrix}.$$

Выразим из (14) вектор искомых переменных $\Sigma$:

$$\Sigma = \Psi^{-1} \cdot Z = \frac{1}{\det \Psi} \cdot \text{adj}(\Psi) \cdot Z. \qquad (15)$$

Умножая обе стороны уравнения (15) на определитель матрицы $\Psi$, получим выражение:

$$\Delta(t)\Sigma = Y(t), \qquad (16)$$

где $\Delta(t) = \det \Psi$ и $Y(t) = \text{adj}(\Psi) \cdot Z$.

Из (16) имеем:

$$\Delta \beta_2 = Y_1 \text{ и } \Delta \beta\omega = Y_2, \qquad (17)$$

где $\beta_2 = \beta^2$ – вспомогательный параметр.

Построим устройство оценки для параметра $\beta_2$ с некоторым $\gamma > 0$:

$$\dot{\hat{\beta}}_2 = \gamma\Delta(Y_1 - \Delta\hat{\beta}_2). \qquad (18)$$

Для невязки $\tilde{\beta}_2$ имеем:

$$\dot{\tilde{\beta}}_2 = -\gamma\Delta^2\tilde{\beta}_2$$

и, как следствие,

$$\tilde{\beta}_2(t) = e^{-\gamma\int_0^t \Delta^2(s)ds} \cdot \tilde{\beta}_2(0). \qquad (19)$$

Обозначим:

$$w = e^{-\gamma\int_0^t \Delta^2(s)ds}. \qquad (20)$$

Заметим, что сигнал $w$ может быть получен с помощью генератора:

$$\dot{w} = -\gamma \cdot \Delta^2 \cdot w, \; w(0) = 1. \qquad (21)$$

Перепишем (19) с учетом (20):

$$\beta_2 - \hat{\beta}_2(t) = w \cdot \left(\beta_2 - \hat{\beta}_2(0)\right). \qquad (22)$$

Из уравнения (22) выразим оценку параметра $\beta_2$ за конечное время (Finite Time или FT):

$$\hat{\beta}_2^{FT} = \beta_2 = \frac{\hat{\beta}_2(t) - \hat{\beta}_2(0)\cdot w}{1-w}. \qquad (23)$$

Для определения знака $\beta$ воспользуемся соотношениями (2) и (17) и получим $\text{sign}(Y_2) = \text{sign}(\Delta\beta)$. Откуда следует, что:

$$\text{sign}(\beta) = \text{sign}(\Delta) \cdot \text{sign}(Y_2). \qquad (24)$$

С учетом (23) и (24) вычислим оценку параметра $\beta$:

$$\widehat{\beta} = \sqrt{\widehat{\beta}_2^{FT} \cdot \text{sign}(\Delta) \cdot \text{sign}(Y_2)}.$$

Построим наблюдатель частоты на основе (17):

$$\dot{\widehat{\omega}} = \widehat{\beta} + \gamma_2 \Delta \widehat{\beta}(Y_2 - \Delta \widehat{\beta}\widehat{\omega}). \qquad (25)$$

Аналогично (23) можно получить оценку частоты за конечное время, модифицируя (25) к виду

$$\widehat{\omega}^{FT}(t) = \frac{\widehat{\omega}(t) - w_1\widehat{\omega}(0) - w_2}{1 - w_1}$$

с использованием вспомогательных переменных $w_1$ и $w_2$

$$\dot{w}_1 = -\gamma_2 \Delta^2 \widehat{\beta} \cdot w_1, w_1(0) = 1.$$

$$\dot{w}_2 = -\gamma_2 \Delta^2 \widehat{\beta}^2 w_2 + w_1, w_2(0) = 0.$$

### Результаты моделирования

Рассмотрим сигнал вида $y(t) = 2\sin(\omega(t) \cdot t + 1)$, $\dot{\omega} = 0{,}05$. Будем использовать фильтры для параметризации (6) с параметрами $\lambda_{1,2,3} = 1,2,3$ и параметры адаптации $\gamma = \gamma_2 = 10^5$.

На рисунке 1 показана временная диаграмма измеряемого сигнала. На рисунках 2 и 3 соответственно показаны оценки параметра β и искомой переменной частоты ω.

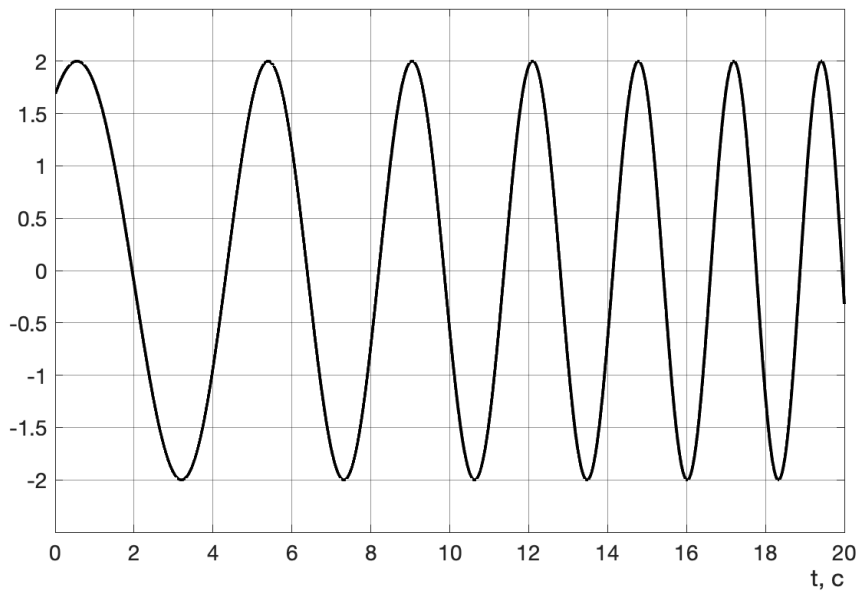

Рис. 1 – Временная диаграмма измеряемого сигнала $y(t)$

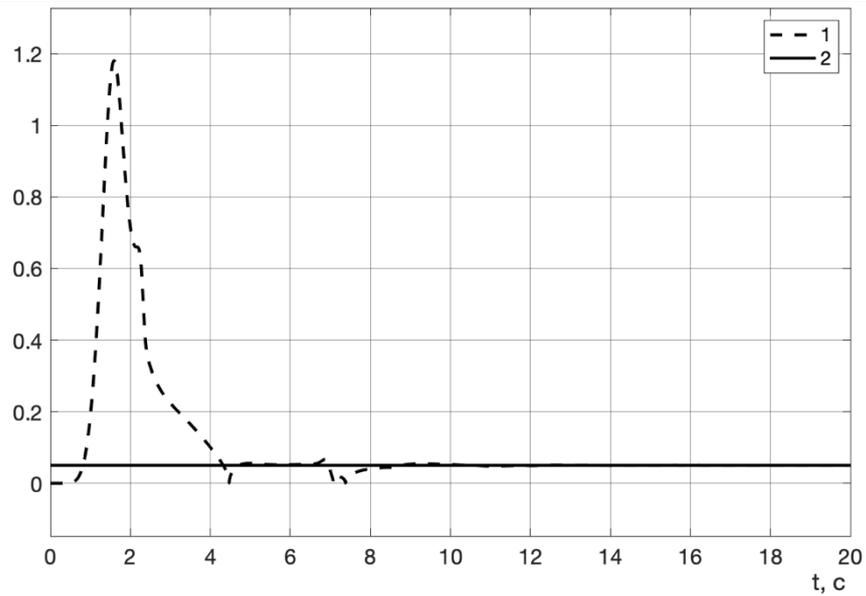

Рис. 2 – Оценка параметра β (1 – оценка, 2 – истинное значение)

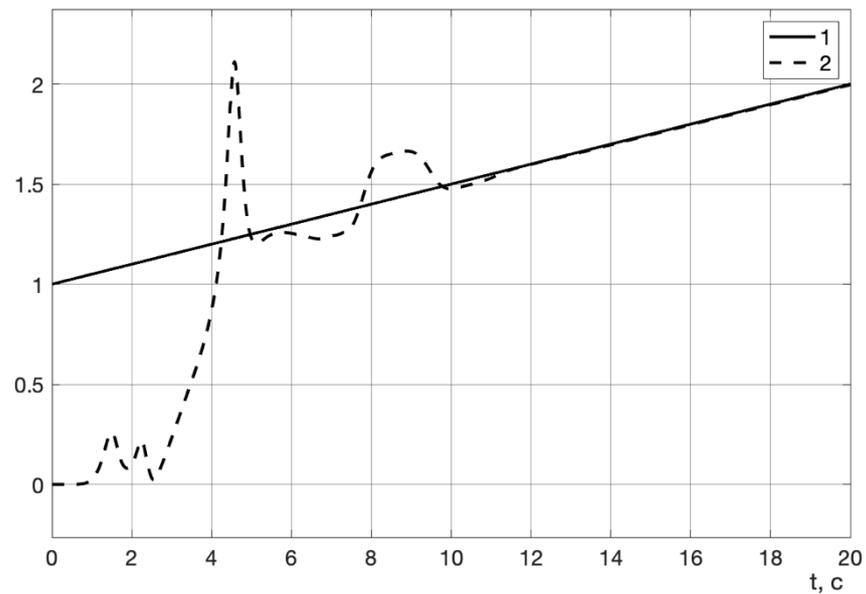

Рис. 3 – Оценка частоты ω (1 – оценка, 2 – истинное значение).

**Заключение**

Был синтезирован алгоритм оценивания переменной частоты синусоидального сигнала. Решение задачи основано на преобразовании модели сигнала к линейному регрессионному уравнению. Задача решена с использованием метода динамического расширения и декомпозиции регрессора (или смешивания регрессора) DREM. Проведено численное моделирование, иллюстрирующее эффективность предложенного алгоритма.

**Приложение**

Воспользуемся Леммой о перестановках:

$$\frac{\lambda}{p+\lambda}xy = x\frac{\lambda}{p+\lambda}y - \frac{1}{p+\lambda}\left(\dot{x}\frac{\lambda}{p+\lambda}y\right) = x\frac{\lambda}{p+\lambda}y - \beta\frac{\lambda}{(p+\lambda)^2}y. \qquad (\text{П.1})$$

Применим фильтр $\frac{\lambda^3}{(p+\lambda)^3}$ к выражению (П.1):

$$\frac{\lambda^3}{(p+\lambda)^3}\dddot{y} = -\beta^2\frac{\lambda^3}{(p+\lambda)^3}t^2\dot{y} - 6\beta^2\frac{\lambda^3}{(p+\lambda)^3}ty - 2\beta\frac{\lambda^3}{(p+\lambda)^3}\omega t\dot{y} -$$

$$-6\beta\frac{\lambda^3}{(p+\lambda)^3}\omega y - \frac{\lambda^3}{(p+\lambda)^3}\omega^2\dot{y}. \qquad (\text{П.2})$$

Перепишем (П.2) с учетом (П.1) для каждого слагаемого:

$$\frac{\lambda^3}{(p+\lambda)^3}t^2\dot{y} = \frac{\lambda^2}{(p+\lambda)^2}\left(t^2\frac{\lambda p}{(p+\lambda)}y - 2\frac{1}{(p+\lambda)}t\frac{\lambda p}{(p+\lambda)}y\right). \qquad (\text{П.3})$$

$$\frac{\lambda^3}{(p+\lambda)^3}\omega t\dot{y} = \omega\frac{\lambda^3}{(p+\lambda)^3}t\dot{y} - 3\beta\frac{\lambda^3}{(p+\lambda)^4}t\dot{y}. \qquad (\text{П.4})$$

$$\frac{\lambda^3}{(p+\lambda)^3}\omega y = \omega\frac{\lambda^3}{(p+\lambda)^3}y - 3\beta\frac{\lambda^3}{(p+\lambda)^4}y. \qquad (\text{П.5})$$

$$\frac{\lambda^3}{(p+\lambda)^3}\omega^2\dot{y} = \frac{\lambda^3}{(p+\lambda)^3}\omega\cdot\omega\cdot\dot{y} = \omega\frac{\lambda^3}{(p+\lambda)^3}\omega\dot{y} - 3\beta\frac{\lambda^3}{(p+\lambda)^4}\omega\dot{y}. \qquad (\text{П.6})$$

Перепишем фильтр из выражения (П.4) с учетом (П.1):

$$\frac{\lambda^3}{(p+\lambda)^3}t\dot{y} = t\frac{\lambda^3 p}{(p+\lambda)^3}y - 3\frac{\lambda^3 p}{(p+\lambda)^4}y. \qquad (\text{П.7})$$

Перепишем фильтр из выражения (П.6) с учетом (П.1):

$$\frac{\lambda^3}{(p+\lambda)^3}\omega\dot{y} = \omega\frac{\lambda^3 p}{(p+\lambda)^3}y - 3\beta\frac{\lambda^3 p}{(p+\lambda)^4}y. \qquad (\text{П.8})$$

Подставим (П.8) в (П.6), (П.7) в (П.4) и полученные выражения вместе с (П.3) и (П.5) подставим в (П.2):

$$\frac{\lambda^3 p^3}{(p+\lambda)^3}y = -\beta^2\left(\frac{\lambda^2}{(p+\lambda)^2}\left(t^2\frac{\lambda p|}{(p+\lambda)}y - 2\frac{1}{(p+\lambda)}t\frac{\lambda p}{(p+\lambda)}y\right)\right) - 6\beta^2\frac{\lambda^3}{(p+\lambda)^3}ty -$$

$$-2\beta\left(\omega\left(t\frac{\lambda^3 p}{(p+\lambda)^3}y - 3\frac{\lambda^3 p}{(p+\lambda)^4}y\right) - 3\beta\frac{1}{(p+\lambda)}\left(t\frac{\lambda^3 p}{(p+\lambda)^3}y - 3\frac{\lambda^3 p}{(p+\lambda)^4}y\right)\right) -$$

$$-6\beta\left(\omega\frac{\lambda^3}{(p+\lambda)^3}y - 3\beta\frac{\lambda^3}{(p+\lambda)^4}y\right) -$$

$$-\left(\omega\left(\omega\frac{\lambda^3 p}{(p+\lambda)^3}y - 3\beta\frac{\lambda^3 p}{(p+\lambda)^4}y\right) - 3\beta\frac{1}{(p+\lambda)}\left(\omega\frac{\lambda^3}{(p+\lambda)^3}y - 3\beta\frac{\lambda^3 p}{(p+\lambda)^4}y\right)\right) =$$

$$= -\beta^2\frac{\lambda^2}{(p+\lambda)^2}t^2\frac{\lambda p}{(p+\lambda)}y + 2\beta^2\frac{\lambda^2}{(p+\lambda)^3}t\frac{\lambda p}{(p+\lambda)}y - 6\beta^2\frac{\lambda^3}{(p+\lambda)^3}ty -$$

$$-2\beta\omega t\frac{\lambda^3 p}{(p+\lambda)^3}y + 6\beta\omega\frac{\lambda^3 p}{(p+\lambda)^4}y + 6\beta^2\frac{1}{(p+\lambda)}t\frac{\lambda^3 p}{(p+\lambda)^3}y - 18\beta^2\frac{\lambda^3 p}{(p+\lambda)^3}y -$$

$$-6\beta\omega\frac{\lambda^3}{(p+\lambda)^3}y + 18\beta^2\frac{\lambda^3}{(p+\lambda)^4}y - \omega^2\frac{\lambda^3 p}{(p+\lambda)^3}y + 3\beta\omega\frac{\lambda^3 p}{(p+\lambda)^4}y -$$

$$-9\beta^2\frac{\lambda^3 p}{(p+\lambda)^5}y + 3\beta\frac{1}{(p+\lambda)}\omega\frac{\lambda^3 p}{(p+\lambda)^3}y. \tag{П.9}$$

Перепишем последнее слагаемое выражения (П.9) с учетом (П.1):

$$\frac{1}{(p+\lambda)}\omega\frac{\lambda^3 p}{(p+\lambda)^3}y = \omega\frac{\lambda^3 p}{(p+\lambda)^4}y - \beta\frac{\lambda^3 p}{(p+\lambda)^5}y. \tag{П.10}$$

Подставим результат (П.10) в (П.9) и получим следующую совокупность фильтров:

$$\frac{\lambda^3 p^3}{(p+\lambda)^3}y = -\beta^2\frac{\lambda^2}{(p+\lambda)^2}t^2\frac{\lambda p}{(p+\lambda)}y + 2\beta^2\frac{\lambda^2}{(p+\lambda)^3}t\frac{\lambda p}{(p+\lambda)}y +$$

$$+6\beta^2\frac{1}{(p+\lambda)}t\frac{\lambda^3 p}{(p+\lambda)^3}y - 6\beta^2\frac{\lambda^3}{(p+\lambda)^3}ty - 6\beta\omega\frac{\lambda^3}{(p+\lambda)^3}y + 18\beta^2\frac{\lambda^3}{(p+\lambda)^4}y -$$

$$-(2\beta\omega t + \omega^2)\frac{\lambda^3 p}{(p+\lambda)^3}y + 12\beta\omega\frac{\lambda^3 p}{(p+\lambda)^4}y - 30\beta^2\frac{\lambda^3 p}{(p+\lambda)^5}y, \tag{П.11}$$

что соответствует выражению (5).

## СПИСОК ЛИТЕРАТУРЫ

1. Besancon, G. Nonlinear observers and applications [Text] / G. Besancon. — Berlin : Springer, 2007. — Vol. 363. — P. 224.

2. Льюнг Л. Идентификация систем: Теория для пользователя. Наука. 1991.

3. Ortega R., Bobtsov A., Pyrkin A., Aranovskiy S. A parameter estimation approach to state observation of nonlinear systems // Systems & Control Letters. 2015. Vol. 85. P. 84-94.

4. Pyrkin A., Bobtsov A., Ortega R., Vedyakov A., Aranovskiy S. Adaptive state observers using dynamic regressor extension and mixing // Systems & Control Letters. 2019. Vol. 133. paper no. 104519.

5. Ortega R., Bobtsov A., Dochain D., Nikolaev N. State observers for reaction systems with improved convergence rates // Journal of Process Control. 2019. Vol. 83. P. 53-62.

6. A.O. Vediakova, A.A. Vedyakov, A.A. Pyrkin, A.A. Bobtsov and V.S. Gromov. The multi-harmonic signal frequencies estimation in finite time // Journal of Physics: Conference Series. 2020. Vol. 1864. DOI:10.1088/1742-6596/1864/1/012116.

7. A.O. Vediakova, A.A. Vedyakov, A.A. Pyrkin, A.A. Bobtsov and V.S. Gromov. Finite Time Frequency Estimation for Multi-Sinusoidal Signals // European Journal of Control. 2021. Vol. 59. P.38-46.

8. Vediakova A., Vedyakov A., Bobtsov A., Pyrkin A. DREM-based Parametric Estimation of Bias-affected Damped Sinusoidal Signals // European Control Conference (ECC 20). 2020. P. 214-219.

9. Во К.Д., Бобцов А.А. Адаптивный наблюдатель переменных состояния линейных нестационарных систем с параметрами, заданными не точно // Автоматика и телемеханика. 2020. №.12. С. 100-110.

10. Ле В.Т., Коротина М.М., Бобцов А.А., Арановский С.В., Во К.Д. Идентификация линейно изменяющихся во времени параметров нестационарных систем // Мехатроника, автоматизация, управление. 2019. Т.20. №.5. С. 259-265.

11. Ван Ц., Ле В.Т., Пыркин А.А., Колюбин С.А., Бобцов А.А. Идентификация кусочно-линейных параметров регрессионных моделей нестационарных детерминированных систем // Автоматика и телемеханика. 2018. №.12. С. 71-82.

12. Данг Х.Б., Пыркин А.А., Бобцов А.А., Ведяков А.А. Синтез адаптивного наблюдателя для нестационарных нелинейных систем с неизвестными полиномиальными параметрами // Научно-технический вестник информационных технологий, механики и оптики. 2021. №3. С. 374-379.

13. Х.Б. Данг, А.А. Пыркин, А.А. Бобцов, А.А. Ведяков. Идентификация полиномиальных параметров нестационарных линейных систем // Известия высших учебных заведений. Приборостроение. 2021. №6. С. 459-468.
14. Sakharov I.I., Shashkin M.A., Nizovtsev S.I. Real-time vibration monitoring // Proceedings of the International Conference on Geotechnics Fundamentals and Applications in Construction: New Materials, Structures, Technologies and Calculations (GFAC 2019). 2019. P. 301-306.
15. Татаркин С.А., Шашкин М.А., Низовцев С.И., Кирпичев А.А., Шкатов П.Н. Исследование нелинейных микродинамических свойств грунта и их применение для целей повышения надежности и эффективности обследования зданий, мониторинга при производстве геотехнических работ в особых геологических условиях // Приборы. 2017. № 3. С. 14-23.

ESTIMATION ALGORITHM NON-STATIONARY FREQUENCY OF THE SINUSOIDAL SIGNAL

S.I. Nizovtsev, S.V. Shavetov, A.A. Pyrkin

**ITMO UNIVERSITY**

Abstract: The article considers the problem of identifying the variable frequency of a sinusoidal signal. To obtain a regression model of the signal, an iterative differentiation of the original analytical expression is performed, and the swapping lemma is applied. The estimation of the parameters of the non-stationary frequency is implemented using the dynamic expansion of the regressor and mixing (DREM) procedure and the Luenberger observer. As a result of the numerical simulation, the efficiency of the proposed algorithm is demonstrated, showing the convergence of the frequency estimate to the true value.

Keywords: parameters identification, non-stationary frequency, the Luenberger observer, the method of dynamically regressor extension and mixing (DREM).